\documentstyle[sprocl]{article}

\input{psfig}

\begin{document}

\title{DARK MATTER CAUSTICS}

\author{P. SIKIVIE, W. KINNEY}

\address{Department of Physics, University of Florida,
Gainesville,\\ FL 32611, USA\\E-mail: sikivie@phys.ufl.edu,
kinney@phys.ufl.edu}

\maketitle\abstracts{ The late infall of cold dark matter onto an 
isolated galaxy such as our own produces flows with definite velocity 
vectors at any physical point in the galactic halo.  It also produces 
caustics which are places where the dark matter density is very 
large.  The outer caustics are topological spheres whereas the inner 
caustics are rings.  The self-similar model of galactic halo formation 
predicts that the caustic ring radii $a_n$ follow the approximate law 
$a_n \sim 1/n$.  In a recent study of 32 extended and well-measured 
galactic rotation curves, we found evidence for this law.}

\section{Introduction}

Before the onset of galaxy formation but after the time $t_{eq}$ of
equality between matter and radiation, the velocity dispersion of the cold
dark matter candidates is very small, of order $\delta v_a (t) \sim 3\cdot
10^{-17} \left( {10^{-5} eV\over m_a}\right)~\left({t_0\over
t}\right)^{2/3}$ for axions and $\delta v_W (t) \sim 10^{-11}
\left({GeV\over m_W}\right)^{1/2}~\left({t_0\over t}\right)^{2/3}$ for
WIMPs, where $t_0$ is the present age of the universe and $m_a$ and $m_W$
are respectively the masses of the axion and the WIMP.  In the context 
of galaxy formation, such small velocity dispersions are entirely 
negligible.  Massive neutrinos, on the other hand, have primordial 
velocity dispersion $\delta v_\nu (t) \simeq 5.3~10^{-4} \left({eV\over
m_\nu}\right)~\left({t_0\over t}\right)^{2/3}$ which is comparable to the
virial velocity in galaxies and therefore non-negligible in the context
of galaxy formation \cite{STG}.  This is the reason why massive neutrinos 
are called `hot dark matter'.

Collisionless dark matter particles lie on a thin 3-dimensional (3D) 
sheet in 6D phase-space.  The thickness of this sheet is the primordial 
velocity dispersion $\delta v$.  If each of the aforementioned species 
of collisionless particles is present, the phase-space sheet has three 
layers, a very thin layer of axions, a medium layer of WIMPs and a thick 
layer of neutrinos.  The phase-space sheet is located on the 
3D hypersurface of points 
$(\vec{r}, \vec{v})~:~\vec{v} = H(t)\vec{r} + \Delta \vec{v} (\vec{r},t)$ 
where $H(t) = {2\over 3t}$ is the Hubble expansion rate and $\Delta \vec{v}
(\vec{r},t)$ is the peculiar velocity field.  Fig. 1 shows a 2D section of
\begin{figure}[t] 
\hfil\psfig{figure=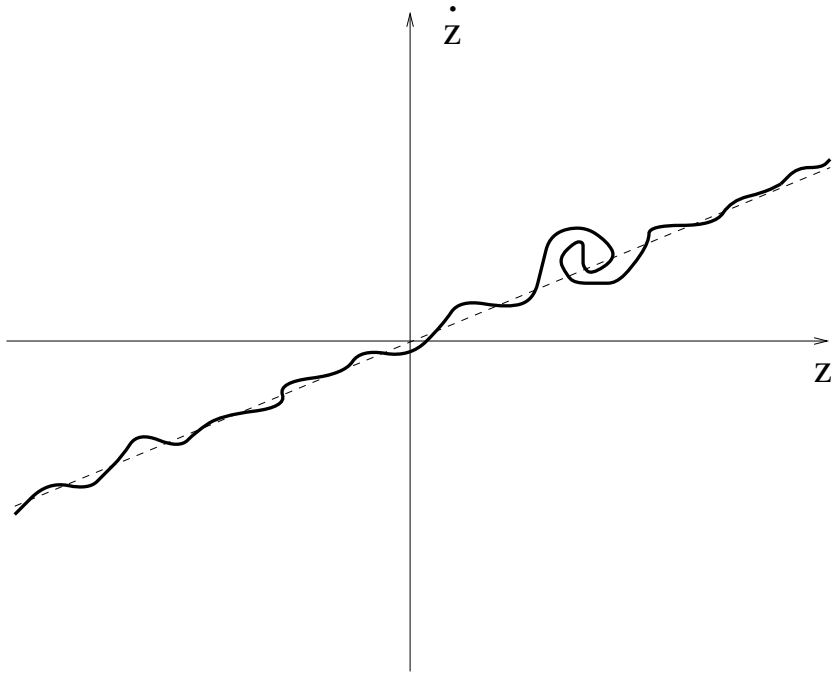,height=3.5in}
\caption{The wiggly line represents the intersection of the $(z,\dot{z})$ 
plane with the 3D sheet on which the collisionless dark matter particles 
lie in phase-space.  The thickness of the line is the primordial velocity
dispersion.  The amplitude of the wiggles in the $\dot{z}$ direction is
the velocity dispersion associated with density perturbations.  Where an
overdensity grows in the non-linear regime, the line winds up in clockwise
fashion.  One such overdensity is shown.}
\end{figure}
6D phase-space along the $(z,\dot{z})$ plane.  The wiggly line is the
intersection of the 3D sheet on which the particles lie in phase-space
with the plane of the figure.  The thickness of the line is the velocity 
dispersion $\delta v$, whereas the amplitude of the wiggles in the line
is the peculiar velocity $\Delta v$.  If there were no peculiar
velocities, the line would be straight since $\dot{z} = H(t) z$ in that
case.

The peculiar velocities are associated with density perturbations and grow
by gravitational instability as $\Delta v\sim t^{2/3}$.  On the other hand
the primordial velocity dispersion decreases on average as $\delta v \sim
t^{-2/3}$, consistently  with Liouville's theorem.  When an overdensity 
enters the non-linear regime, the particles in its vicinity fall back onto 
it.  This implies that the phase-space sheet `winds up' there in clockwise 
fashion.  One such overdensity is shown in Fig. 1.  In the linear regime, 
there is only one value of velocity, i.e. one single flow, at a typical 
location in physical space, because the phase-space sheet covers physical 
space only once.  On the other hand, inside an overdensity in the non-linear 
regime, the phase-space sheet covers physical space multiple times implying 
that there are several (but always an odd number of) flows at such locations.

At the boundary surface between two regions one of which has $n$ flows and 
the other $n + 2$ flows, the physical space density is very large because 
the phase-space sheet has a fold there.  At the fold, the phase-space sheet 
is tangent to velocity space and hence, in the limit of zero velocity   
dispersion $(\delta v = 0)$, the density diverges since it is
the integral of the phase-space density over velocity space.  The
structure associated with such a phase-space fold is called a 'caustic'.   
It is a surface in physical space.  It is easy to show that, in the limit 
of zero velocity dispersion, the density diverges as 
$d \sim {1\over\sqrt{\sigma}}$ when the caustic is approached from the side 
with $n+2$ flows, where $\sigma$ is the distance to the caustic.  Velocity 
dispersion cuts off the divergence.

As mentioned above, the process of galactic halo formation involves the 
local winding up of the phase-space sheet of collisionless dark matter 
particles.  If the galactic center is approached from an arbitrary direction 
at a given time, the local number of flows increases.  First, there is 
one flow, then three flows, then five, seven...  The number of flows at
our location in the Milky Way galaxy today has been estimated \cite{is} to 
be of order 100.  The boundary between the region with one (three, five, 
...) and the region with three (five, seven, ...) flows is the location of 
a caustic which is topologically a sphere surrounding the galaxy.  When
these caustic spheres are approached from the inside the density diverges
as $d\sim {1\over\sqrt{\sigma}}$ in the zero velocity dispersion limit.
These spheres are the outer caustics in the phase-space structure of
galactic halos.  In addition there are inner caustics.

It is a little more difficult to see why there must be inner caustics,  
and to derive their structure.  See refs. [3,4] for details.  The inner
caustics are rings.  They are located near where the particles with the 
most angular momentum in a given in and out flow are at their distance 
of closest approach to the galactic center.  A ring is a closed tube 
whose cross-section is a $D_{-4}$ catastophe \cite{Gilmore}.  The 
cross-section is shown in Fig.~2 in the limit of axial and reflection 
symmetry, and where the transverse dimensions, $p$ and $q$, are much 
smaller than the ring radius $a$.  
\begin{figure}[t]
\hfil\psfig{figure=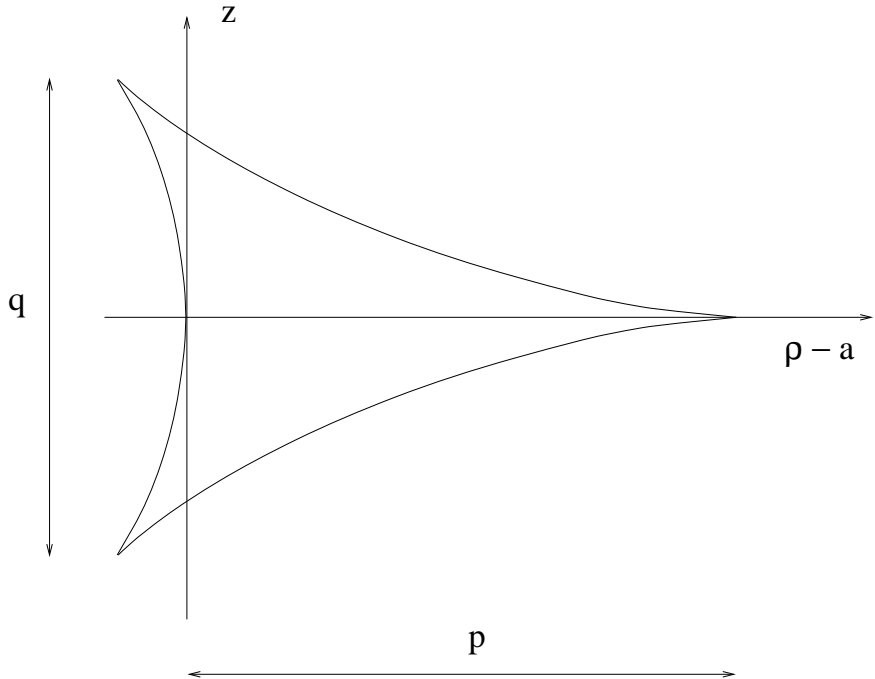,height=3.5in}
\caption{Cross-section of a caustic ring in the case of axial 
and reflection symmetry.  The galactic center is to the left of the
figure.  The $z$-direction is orthogonal to the galactic plane.  The 
$\rho$-direction is radial. $a$ is the caustic ring radius.  The 
closed line with three cusps is the location of a caustic surface. 
The density diverges when the surface is approached from the inside 
as $\sigma^{-1/2}$ where $\sigma$ is the distance to the surface.}
\end{figure}
In the absence of any symmetry, the cross-section of the tube does 
not have the exact shape shown in Fig.~2 but it still has that 
shape qualitatively, i.e. it is still a closed line with three 
cusps one of which points away from the galactic center. 

The existence of caustic rings of dark matter follows from only two
assumptions:
\begin{enumerate}
\item the existence of collisionless dark matter
\item that the velocity dispersion of the infalling dark matter is much    
less, by a factor ten say, than the rotation velocity of the galaxy.    
\end{enumerate}

\noindent Only the second assumption requires elaboration.  Velocity 
dispersion has the effect of smoothing out caustics. The question is
when is the velocity dispersion so large as to smooth caustic rings over
distance scales of order the ring radius $a$, thus making the notion
of caustic ring meaningless.  In ref. [3] this critical velocity
dispersion was estimated to be 30 km/s = $10^{-4}$ for the caustic rings
in our own galaxy, whose rotation velocity is 220 km/s.  $10^{-4}$ is much 
less than the {\it primordial} velocity dispersion $\delta v$ of the cold
dark matter candidates.  However the velocity dispersion $\Delta v$
associated with density perturbations also smoothes caustics in coarse
grained observations.  So the question is whether the velocity  
dispersion $\Delta v$ of cold dark matter particles associated with 
density perturbations falling onto our galaxy is less than 30 km/s.  The
answer is yes with high probability since the infalling dark matter
particles are not associated with any observed inhomogeneities.  30 km/s 
is of order the velocity dispersion of the Magellanic Clouds.  For the 
velocity dispersion of the dark matter particles presently falling onto 
our galaxy to be as large 30 km/s, these particles would have to be part 
of clumps whose mass/size ratio is of order that of the Magellanic Clouds.  
But if that were the case, why did these clumps fail to become luminous?

One might ask whether caustic rings can be seen in N-body simulations
of galaxy formation.  The generic surface caustics associated with 
simple folds of the phase-space sheet have been seen \cite{Melott}. 
However, caustic rings would require far greater resolution than    
presently available, at least in a 3D simulation of our own halo.
Indeed, the largest ring in our galaxy has been estimated \cite{me} 
to have radius of order 40 kpc.  It is part of an in and out flow 
that extends to the Galaxy's current turnaround radius, of order 2 Mpc.  
To resolve this first ring, the spatial resolution would have to be
considerably smaller than 10 kpc.  Hence a minimum of
$2\cdot{1 \over (10{\rm kpc})^3} {4\pi \over 3}(2 {\rm Mpc})^3
\simeq 7\cdot 10^7$ particles would be required to see the caustic   
ring in a simulation of this one flow.  However, the number of flows
at 40 kpc in our halo \cite{sty} is of order 10.  So it appears
that $10^9$ particles is a strict minimum in a 3D simulation of
our halo.  Even so, this addresses only the kinematic requirement of
resolving the halo in phase-space, assuming moreover that the particles
are approximately uniformly distributed on the phase-space sheets.  There
is a further dynamical requirement that 2-body collisions do not
artificially 'fuzz up' the phase-space sheets.  Indeed 2-body collisions
are entirely negligible in the flow of cold dark matter particles such as
axions or WIMPs.  On the other hand, 2-body collisions are present, and
hence the velocity dispersion is artificially increased, in the 
simulations.  This may occur to such an extent that the caustics are 
washed away even if $10^9$ particles are used. 

In the self-similar infall model \cite{ss,sty} of galactic halo 
formation the caustic ring radii $a_n$ are predicted \cite{me}:
\begin{equation}
\{a_n: n=1,2,...\} \simeq (39,~19.5,~13,~10,~8,...) {\rm kpc}   
\left({j_{\rm max}\over 0.25}\right) \left({0.7\over h}\right)
\left({v_{rot} \over 220 {km\over s}} \right)
\label{rad}
\end{equation}
where $h$ is the present Hubble rate in units of 100 km/s.Mpc, $v_{rot}$
is the rotation velocity of the galaxy and $j_{\rm max}$ is the maximum 
of its dimensionless angular momentum distribution as defined in 
ref. [8].  In Eq.~(1) we assume that the parameter \cite{ss,sty} 
$\epsilon = 0.3$. 

Eq.~(\ref{rad}) predicts the caustic ring radii of a galaxy in terms of 
its first ring radius $a_1$.  If the caustic rings lie close to the 
galactic plane they cause bumps in the rotation curve, at the caustic 
ring radii.  As a possible example of this effect, consider \cite{me} 
the rotation curve of NGC3198, one of the best measured.  It has three 
faint bumps at radii: 28, 13.5 and 9 kpc, assuming $h=0.75$.  The ratios 
happen to be consistent with Eq.~(\ref{rad}) assuming the bumps are caused 
by the first three ($n=1,2,3$) ring caustics of NGC3198.  Moreover, since 
$v_{\rm rot} = 150$ km/s, $j_{\rm max}$ is determined to equal 0.28.  
The uncertainty in $h$ is a systematic effect that can be corrected for 
when determining $j_{\rm max}$ because the bump radii scale like 
$1/h^\prime$ where $h^\prime$ is the Hubble rate assumed by the 
observer in constructing the rotation curve, and the caustic ring radii 
scale as $1/h$.  Rises in the inner rotation curve of the Milky Way were 
also interpreted \cite{me} as due to caustics $n =6,7,8,9,10,11,12$ and 13.  
This determined the value of $j_{\rm max}$ of our own galaxy to be 0.263.  
The first five caustic ring radii in our galaxy are then predicted to be: 
41,~20,~13.3,~10,~8~kpc.

\section{Evidence for universal structure in galactic halos}

Motivated by these findings, we analyzed \cite{fea} a set of 32 extended 
well-measured galactic rotation curves which had been previously selected 
\cite{kbrh} under the criteria that each is an accurate tracer of the 
galactic radial force law, and that it extends far beyond the edge of 
the luminous disk.

According to the self-similar caustic ring model, each galaxy has its own   
value of $j_{\rm max}$.  Over the set of 32 galaxies selected in ref. [10], 
$j_{\rm max}$ has some unknown distribution.  However, the fact that the 
values of $j_{\rm max}$ of NGC3198 and of the Milky Way happen to be close 
to one another, within 7\%, suggests that the $j_{\rm max}$ distribution 
may be peaked near a value of 0.27~.  Our strategy is to rescale each 
rotation curve according to
\begin{equation}
r \rightarrow \tilde r  = r~({220\,{\rm km/s} \over v_{rot}})
\label{rescale}
\end{equation}
and to add them in some way.  Since Eq.~(\ref{rad}) 
predicts the $n$th caustic radius $a_n$ to be distributed like $j_{\rm max}$ 
for all n, and it fixes the ratios $a_n/a_1 \simeq 1/n$, the sum of rotation 
curves should show the $j_{\rm max}$ distribution, once for $n=1$, then at 
about half the $n=1$ radii for $n=2$, then at about 1/3 the $n=1$ radii for 
$n=3$,and so on.  If the $j_{\rm max}$ distribution is broad, the sum of 
rotation curves is unlikely to show any feature.  However, if it is peaked, 
then the sum should show a peak for $n=1$ at some radius, then again at 1/2 
that radius for $n=2$, at 1/3 the radius for $n=3$, and so on.  If the 
$j_{\rm max}$ distribution is peaked at 0.263 (the value for the Milky Way)
the peaks in the sum of rotation curves should appear at 41 kpc, 20 kpc, 
13.3 kpc ...
 
The procedure followed to add the 32 rotation curves is described in detail
in ref. [9].  Briefly, we proceeded as follows.  For each rotation 
curve, all data points with rescaled radii $\tilde r <10\,{\rm kpc}$ were 
deleted to remove the effect of the luminous disk.  The remaining points  
were then fitted to a line.  The rotation velocity $v_{rot}$ used to rescale 
the radii in Eq. (\ref{rescale}) is the average of that line.  The rms 
deviation $\sqrt{\left\langle \delta v~^2 \right\rangle}$ from the linear 
fit was determined for each galaxy.  This was taken to be the error on the 
residuals $\delta v_i$, i.e. the differences between the measured velocities 
in a rotation curve and the linear fit.  Finally the sample of 32 galaxies 
was averaged in $2\,{\rm kpc}$ radial bins:
\begin{equation}
b_i \equiv {1 \over N_i} \sum_{j=1}^{N_i} \delta \tilde v_j,
\end{equation}
where $N_i$ is the number of data points in the bin. The assigned error on   
each $b_i$ is then simply $1 /\sqrt{N_i}$.  Fig.~3 shows the result.
\begin{figure}[t]
\hfil\psfig{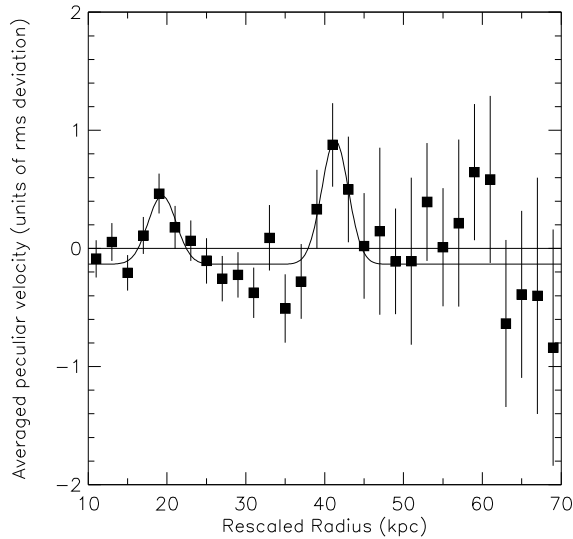}
\caption{Binned data for 32 galaxy sample, with peaks fit to Gaussians.}
\end{figure}

There are two features evident at roughly $20$ and $40\,{\rm kpc}$.  A fit
to two Gaussians plus a constant indicates features at 
$19.4 \pm 0.7\,{\rm kpc}$ and $41.3 \pm 0.8\,{\rm kpc}$, with overall 
significance of $2.4\sigma$ and $2.6\sigma$, respectively.  Fig.~3 shows 
the fitted curve.  When the same fit is applied to the same data in 
1 kpc bins, the significance of the two peaks is 2.6  and 3.0 $\sigma$ 
respectively.  The locations of the features agrees with the predictions 
of the self-similar caustic ring model with the $j_{\rm max}$ distribution 
peaked at 0.27.  The use of Gaussians to fit the peaks in the combined 
rotation curve was an arbitrary choice in the absence of information 
on the $j_{\rm max}$ distribution.

The existence of velocity peaks and caustic rings in the cold dark 
matter distribution is relevant to axion \cite{Bux} and WIMP searches
\cite{Krauss}.  Caustics may also be investigated using gravitational 
lensing techniques \cite{Hogan}. 

\section*{Acknowledgements:}

This work was supported in part by the US Department of Energy under 
grant No. DEFG05-86ER40272.

\end{document}